\documentclass[conference]{IEEEtran}
\IEEEoverridecommandlockouts
\usepackage{cite}
\usepackage{amsmath,amssymb,amsfonts}
\usepackage{algorithmic}
\usepackage{graphicx}
\usepackage{textcomp}
\usepackage{xcolor}
\usepackage{url}
\usepackage{tabularx}
\usepackage{booktabs} % Indispensable pour de superbes lignes horizontales académiques
\usepackage{array}    % Permet de contrôler finement l'alignement des colonnes

\usepackage{cite}
\usepackage[table]{xcolor}
\usepackage[most]{tcolorbox}
\definecolor{myblue}{RGB}{235, 245, 255}

\newcolumntype{L}{>{\raggedright\arraybackslash}X}

\def\BibTeX{{\rm B\kern-.05em{\sc i\kern-.025em b}\kern-.08em
    T\kern-.1667em\lower.7ex\hbox{E}\kern-.125emX}}
\begin{document}

%\title{Open WebXR or Commercial Game Engines for the Metaverse? A Comparative Analysis of Technical, Economic, Ethical, and Sustainability Trade-offs
%\title{Open WebXR versus Commercial Game Engines: Implications for an Open, Sustainable, and Interoperable Metaverse
\title{WebXR and Commercial Game Engines for the Metaverse: A Socio-Technical Analysis of Openness, Interoperability, and Sustainability
\thanks{This work was supported by the MUSMET project funded by the EIC Pathfinder Open scheme of the European Union (grant agreement n. 101184379). Views and opinions expressed are however those of the authors only and do not necessarily reflect those of the European Union or the European Innovation Council. Neither the European Union nor the European Innovation Council can be held responsible for them.}
}

\author{\IEEEauthorblockN{Luca Turchet}
\IEEEauthorblockA{\textit{Department of Information Engineering} \\
\textit{and Computer Science} \\
\textit{University of Trento}\\
Trento, Italy \\
luca.turchet@unitn.it}
\and
\IEEEauthorblockN{Michel Buffa}
\IEEEauthorblockA{\textit{University Côte d'Azur} \\
\textit{SPARKS research group}\\
I3S Laboratory\\
Sophia-Antipolis, France \\
michel.buffa@univ-cotedazur.fr}
%\and
}

\maketitle

\begin{abstract}
The Metaverse is often framed as a persistent, interoperable, and embodied network of virtual and augmented environments. Yet, most contemporary XR applications are developed through commercial game engines and distributed through proprietary app stores, creating tensions between openness and platform dependency. This paper critically examines open WebXR technologies with conventional commercial game-engine pipelines, with particular attention to XR hardware, software architectures, developer workflows, governance, ethics, interoperability, and sustainability. We argue that WebXR may provide a viable route toward a more accessible, device-independent, and institutionally sustainable Metaverse, especially for education, research, cultural heritage, prototyping, and public-interest applications. At the same time, commercial engines remain advantageous for graphically intensive, low-latency, deeply integrated, and large-scale XR products. The paper concludes that the choice should not be framed as WebXR versus engines, but as a continuum: WebXR is preferable when accessibility, interoperability, low-friction deployment, and long-term maintainability are primary goals, whereas native engines remain preferable when performance, platform-specific hardware access, and production-grade tooling dominate.
\end{abstract}


% ci sono stati altri che hanno fatto questo tipo di studio o considerazione? CERCA lavori su WebXR.....

%%%%%%%%%%%%%%%%%%%%%%%%%%%%%%%%%%%%%%%%%%%%%%%%%%%%%%%%%%%%%%%%
\section{Introduction} 
\label{sec:intro}

The Metaverse has been described as an emerging socio-technical ecosystem of persistent, shared, and immersive digital environments \cite{wang2022survey}. Although the term is relatively recent in its current industrial and academic usage, its realization depends on long-standing technical questions: how virtual worlds are rendered, how users access them, how identity and assets move across platforms, and who controls the infrastructures through which participation occurs. In practice, much Metaverse development through Extended Reality (XR) technologies currently relies on commercial game engines such as Unity and Unreal Engine, combined with proprietary distribution channels such as the Apple App Store, Google Play, Meta Quest Store, Steam, or enterprise device-management systems.

This model has clear advantages. Commercial engines provide mature tooling, advanced rendering pipelines, asset stores, physics, animation systems, profiling tools, and native access to device-specific XR features. However, it also introduces dependencies on platform owners, app-store rules, engine licensing, device accounts, operating-system updates, and hardware support cycles. These dependencies are particularly relevant for universities, cultural institutions, small studios, civic projects, and developers seeking alternatives to ecosystems dominated by large US technology companies.

WebXR offers a different paradigm. Instead of installing a native application, users access immersive content through a web browser \cite{maclntyre2018thoughts}. The WebXR Device API is currently a W3C Candidate Recommendation Draft intended to become a W3C Recommendation, but it is still formally a work in progress rather than a completed standard\footnote{\url{https://www.w3.org/TR/webxr/}}. WebXR builds on the broader Web platform and can be combined with other W3C APIs such as WebGL \cite{congote2011interactive}, WebGPU \cite{rodriguez2025cross}, WebAudio \cite{smus2013webaudio}, WebRTC \cite{sredojev2015webrtc}, WebAssembly \cite{zhang2025research}, Web Components \cite{yang2002web}, with higher level open  source libraries like Three.js \cite{danchilla2012three}, Babylon.js \cite{moreau2016babylon}, A-Frame \cite{santos2019web}, and networked web frameworks. Recent work has argued that WebXR can support open, cross-platform, web-based Metaverse architectures \cite{macario2024open, buffa2025interactive}, while comparative work suggests that native engines retain advantages in some areas but that WebXR is increasingly viable for many immersive analytics and research applications~\cite{butcher2024native}.%-> controllare referenze questa ultima pagina

However, previous comparisons have primarily focused on technical development aspects, such as implementation effort, rendering capabilities, or immersive analytics workflows. In contrast, this paper adopts a broader socio-technical perspective that explicitly considers governance, openness, standards, developer autonomy, multimodal interaction (including audio and haptics), sustainability, institutional deployment, and long-term interoperability. Rather than evaluating development platforms solely from a software engineering perspective, we examine their implications for the future evolution of the Metaverse as an open socio-technical ecosystem.

Despite growing industrial interest in the Metaverse, a fundamental tension remains largely unresolved: whether future immersive ecosystems should be built primarily on proprietary platform infrastructures or on open web technologies. Much of today's XR ecosystem is controlled by a small number of technology companies that simultaneously govern hardware platforms, operating systems, application stores, identity systems, and monetization channels. As a result, developers, researchers, educational institutions, and public organizations increasingly face questions concerning platform lock-in, long-term sustainability, governance, interoperability, and digital sovereignty.

These concerns are particularly relevant in Europe, where policy initiatives such as the Digital Markets Act, the European Declaration on Digital Rights and Principles, and the European Commission's initiatives on Web 4.0 and Virtual Worlds emphasize openness, interoperability, technological sovereignty, competition, and user empowerment. In this context, the technological foundations of the Metaverse are increasingly viewed not merely as engineering choices but as strategic decisions that may influence the future governance and accessibility of immersive digital ecosystems.
While commercial game engines have been extensively studied from technical and production perspectives, comparatively little attention has been devoted to understanding how the younger WebXR proposal may contribute to a more open and sustainable Metaverse ecosystem. Existing discussions often focus on rendering performance or usability considerations \cite{yu2023survey,rzeszewski2021usability}, whereas broader implications related to governance, developer autonomy, software longevity, educational deployment, and ethical concerns remain fragmented across different research communities \cite{rodriguez2021democratizing,hanfati2022design,marti2023using}.

This paper adopts the form of a position paper grounded in a qualitative socio-technical comparative analysis. Rather than conducting controlled benchmarking experiments, we synthesize technical documentation, standards specifications, prior empirical studies, industrial practices, and recent literature to examine the relative strengths, limitations, and long-term implications of WebXR and commercial XR ecosystems. The analysis is structured around socio-technical systems theory \cite{baxter2011socio}, treating XR ecosystems as configurations of technologies, institutions, workflows, governance models, and deployment practices. Our objective is not to provide definitive performance measurements, but rather to articulate an evidence-informed perspective on the role of WebXR within a more open, sustainable, and interoperable Metaverse ecosystem.

The comparative analysis was informed through a structured review of four categories of sources: (i) official standards and technical specifications (e.g., W3C WebXR and related Web APIs, OpenXR), (ii) peer-reviewed scientific literature on WebXR, XR development, and Metaverse technologies, (iii) publicly available technical documentation and developer guidelines from major XR platforms and engine vendors, and (iv) documented industrial practices concerning deployment, distribution, licensing, and ecosystem governance. Sources were selected based on their relevance to at least one of the socio-technical dimensions considered in this work, namely hardware compatibility, software architecture, multimodal interaction, deployment, governance, developer accessibility, economic factors, ethics, or sustainability. Evidence extracted from these sources was subsequently synthesized qualitatively rather than quantitatively in order to compare WebXR and commercial XR ecosystems across common analytical dimensions.

In addition, the comparison between WebXR and native XR platforms should not be restricted to visual rendering alone. Today's Metaverse experiences increasingly rely on multimodal interaction combining graphics, spatial audio, haptics, embodiment, and social communication. While graphics performance has traditionally dominated discussions surrounding WebXR adoption, the relative maturity of web-based audio technologies and the more limited support for advanced haptic devices introduce additional dimensions that must be considered when evaluating the suitability of WebXR for future immersive ecosystems.
%
%Recent research has also argued that audio remains a less explored dimension of Metaverse infrastructures. Boem et al.~\cite{boem2025issues} note that many discussions of the Metaverse emphasize visual immersion and networking while overlooking the technological requirements associated with real-time audio communication, spatial audio rendering, and collaborative creative activities. This observation further motivates the need for a broader comparison that considers multimodal interaction capabilities alongside traditional rendering and deployment concerns.

This paper addresses the following research questions:

\begin{itemize}
\item RQ1: What are the main technical advantages and limitations of WebXR compared to commercial game-engine approaches for Metaverse development?

\item RQ2: How do WebXR and commercial game-engine ecosystems differ with respect to interoperability, governance, developer autonomy, and sustainability?

\item RQ3: Under which application scenarios is WebXR a preferable solution, and when do native game engines remain the most appropriate choice?
\end{itemize}

%This paper provides a structured comparison between WebXR and commercial game-engine approaches for Metaverse development. The central claim is that WebXR should be considered not merely as a lightweight alternative to native XR, but as a governance and sustainability strategy for an open Metaverse.

% Identify gaps.... 

% technical gap: latency... !! -> citare paper Alberto su Issues.... -> lì si parlava di webXR : Rileggere

%%%%%%%%%%%%%%%%%%%%%%%%%%%%%%%%%%%%%%%%%%%%%%%%%%%%%%%%%%%%%%%%
\section{Background}

\subsection{Commercial Game-Engine Pipelines}
Commercial XR pipelines typically involve a game engine, platform SDKs, build targets, app-store submission, and device-specific testing. Unity and Unreal Engine dominate many XR workflows because they provide high-level abstractions for rendering, input, physics, shaders, animation, spatial audio, and deployment. These engines are especially strong when developers require high visual fidelity, optimized native performance, access to platform-specific SDKs, or integration with commercial content pipelines. 

Other engines also contribute to the XR ecosystem. In particular, the open-source Godot Engine has gained increasing adoption owing to its permissive licensing model, active community, and growing XR support through OpenXR. Although its ecosystem and tooling are currently less mature than those of Unity or Unreal Engine for many professional XR applications, Godot represents an important intermediate point between fully proprietary commercial engines and browser-native WebXR development, illustrating that openness can also be pursued within native-engine ecosystems.

However, the benefits of game engines come with dependencies. Developers must track engine-version changes, licensing terms, build-system updates, XR SDK changes, platform review rules, and store policies. Unity's 2023--2024 pricing controversy, followed by the cancellation of its Runtime Fee in 2024, illustrated how engine-governance decisions can create uncertainty for developers\footnote{\url{https://unity.com/products/pricing-updates}}. Unreal Engine currently applies a 5\% royalty on lifetime gross revenue above \$1 million for products directly attributable to Unreal Engine, unless specific exemptions apply\footnote{\url{https://www.unrealengine.com/license}}. These models may be acceptable for commercial studios, but they matter for smaller organizations and public-interest projects. 
Furthermore, native distribution paradigms expose logistical hurdles when deployed in non-commercial settings. Commercial VR headsets operate heavily under a ``smartphone'' model, which is intrinsically linked to a single consumer account and attached to a single user. This architecture is often poorly suited for scenarios common to the Cultural and Creative Industries, such as museums, public exhibitions, or educational student projects. In these environments, hardware may be shared among a large set of rotating users, making the mandatory application installation and manual store updates across a fleet of proprietary devices a significant operational barrier.

\subsection{Open WebXR Pipelines}

\begin{figure}[ht!]
\centerline{\includegraphics[width=0.74\columnwidth]{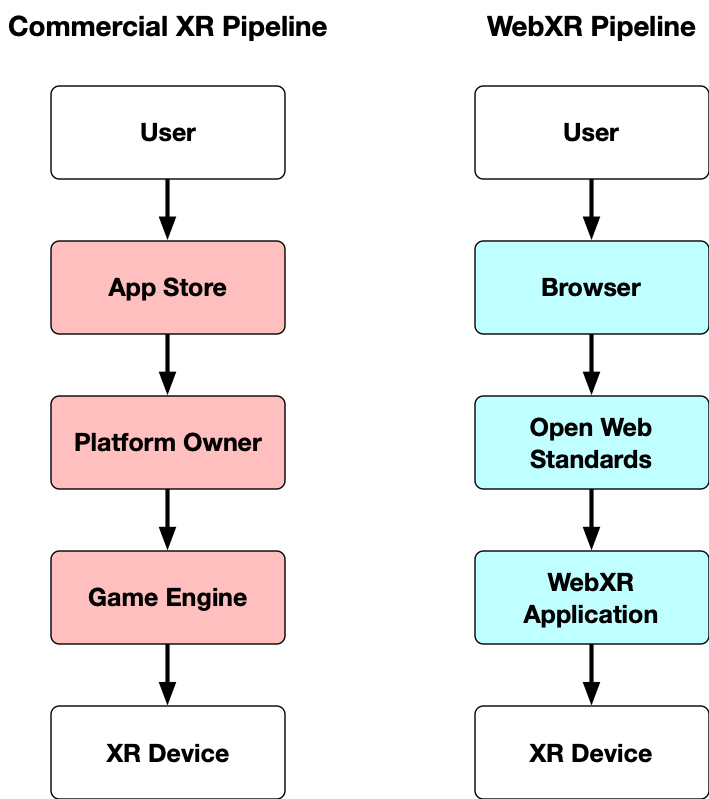}}
\caption{Comparison between conventional XR application ecosystems and WebXR-based deployment architectures. 
%While commercial pipelines rely on multiple centralized intermediaries (app stores, platform owners, engine vendors), WebXR leverages open web standards and browser-based access, reducing deployment friction and platform dependence.
}
\label{fig:pipelines}
\end{figure}

WebXR shifts XR deployment from compiled native applications to browser-accessible experiences delivered through URLs. Rather than requiring application installation, immersive content can be hosted on standard web infrastructure and accessed through compatible browsers and XR devices. This model reduces deployment friction, simplifies updates, and may lower barriers to access in educational, research, and public-facing contexts. The Immersive Web community describes WebXR as supporting deployment across XR platforms through browser-based runtimes\footnote{\url{https://developer.mozilla.org/en-US/docs/Web/API/WebXR_Device_API}}, while the WebXR Device API enables rendering and interaction with immersive hardware through standardized interfaces \cite{maclntyre2018thoughts}, as shown in Fig.~\ref{fig:pipelines}.

From an architectural perspective, WebXR builds upon the broader Web ecosystem and can be combined with complementary technologies including WebGL, WebGPU, WebAudio, WebRTC, WebAssembly, and open asset formats such as glTF \cite{congote2011interactive,rodriguez2025cross,smus2013webaudio,sredojev2015webrtc, zhang2025research}. These technologies support rendering, networking, audio processing, communication, and computation within browser environments, thereby enabling interoperable XR applications grounded in open standards. Such characteristics align with broader interoperability efforts, including those promoted by the Metaverse Standards Forum.%\footnote{\url{https://metaverse-standards.org}}.

At the same time, WebXR development remains subject to challenges associated with evolving standards and heterogeneous browser support. Several relevant APIs are still under active standardization and may exhibit inconsistent implementation across browsers and XR devices. Previous work has reported practical difficulties including fragmented browser support, testing complexity, maintenance overhead, and adaptation to evolving implementations~\cite{zubair2021long}.

To mitigate these issues, developers frequently rely on higher-level frameworks such as Three.js, Babylon.js, and A-Frame, which abstract lower-level browser APIs and provide reusable components for rendering, interaction, and XR scene management. These frameworks can reduce implementation complexity and facilitate development across heterogeneous devices.

%%%%%%%%%%%%%%%%%%%%%%%%%%%%%%%%%%%%%%%%%%%%%%%%%%%%%%%%%%%%%%%%%%%%%%%%%%%%%%%%
\section{Comparative Analysis}

Rather than focusing exclusively on rendering performance or software engineering considerations, our comparison adopts a broader socio-technical perspective encompassing hardware compatibility, software architectures, multimodal interaction capabilities, deployment workflows, developer accessibility, economic factors, governance models, ethical implications, and sustainability. These dimensions are examined individually in the following subsections and subsequently synthesized in Table~\ref{tab:comparison}.
%and Figure~\ref{fig:radar_comparison}.

\begin{table*}[t]
\caption{Comparison between WebXR and Commercial Game-Engine Approaches for Metaverse Development}
\label{tab:comparison}
\centering
\renewcommand{\arraystretch}{1.3}
\begin{tabular}{p{3.2cm}p{6.5cm}p{6.5cm}}
\hline
\textbf{Dimension} & \textbf{WebXR} & \textbf{Commercial Game Engines} \\
\hline

\rowcolor{myblue}
\textbf{Deployment} &
Access through URLs; no installation required; updates are immediate and server-side. &
Requires application installation, app-store distribution or sideloading, and client-side updates. \\

\textbf{Hardware Independence} &
Runs across compatible browsers and devices; one codebase can target multiple XR platforms. &
Typically requires platform-specific builds and testing for each target device. \\

\rowcolor{myblue}
Performance &
Good for lightweight-to-medium complexity applications; performance constrained by browser environment. &
Generally superior rendering, optimization, and low-level hardware access. \\

\textbf{Access to Device Features} &
Dependent on browser support for XR APIs. &
Direct access to vendor SDKs and advanced hardware features. \\

\rowcolor{myblue}
Development Skills &
Leverages web-development skills (HTML, JavaScript, TypeScript). &
Requires engine-specific expertise and workflows. \\

\textbf{Distribution Control} &
Developers can self-host and distribute experiences independently. &
Dependent on platform stores and approval processes. \\

\rowcolor{myblue}
\textbf{Monetization} &
Flexible business models with fewer platform restrictions. &
Often subject to platform fees and store policies. \\

\textbf{Interoperability} &
Built on open web standards (HTTP, WebRTC, WebAudio, WebGPU, glTF). &
Can support standards but often involves proprietary project formats and dependencies. \\

\rowcolor{myblue}
\textbf{Maintenance} &
Centralized updates; users always access the latest version. &
Requires maintaining multiple application builds and managing version compatibility. \\

\textbf{Educational Use} &
Particularly suitable for classrooms, laboratories, museums, and public institutions. &
Device provisioning and installation management can be cumbersome. \\

\rowcolor{myblue}
\textbf{Governance} &
Aligned with open-web principles and decentralized deployment. &
Controlled by engine vendors, platform providers, and app-store operators. \\

\textbf{Security} &
No centralized review process; security responsibilities lie largely with developers. &
Benefits from store-based review, certification, and platform-level controls. \\

\rowcolor{myblue}
\textbf{Sustainability} &
Reduced software obsolescence through web standards and browser compatibility. &
Can suffer from store discontinuation, SDK deprecation, and platform lock-in. \\

\textbf{Graphical Fidelity} &
Improving rapidly but still limited compared to native rendering pipelines. &
State-of-the-art visual quality and optimization. \\

\rowcolor{myblue}
\textbf{Audio Support} &
Strong support through Web Audio API and WebRTC; suitable for spatial audio, communication, and collaborative applications. &
Advanced audio pipelines, middleware integration (FMOD, Wwise), acoustic simulation, and professional audio workflows. \\

\textbf{Haptic Support} &
Limited and browser-dependent support, typically restricted to basic controller vibration. &
Broad support for advanced haptic devices, wearable interfaces, force-feedback systems, and vendor SDKs. \\

\rowcolor{myblue}
\textbf{Best Suited For} &
Education, research, cultural heritage, social XR, collaborative environments, rapid prototyping. &
AAA games, industrial simulations, digital twins, high-end commercial products. \\

\hline
\end{tabular}
\end{table*}

% HERE RADAR CHART

%\begin{figure}[ht!]
%\centerline{\includegraphics[width=\columnwidth]{./Figures/radar_chart}}
%\caption{Qualitative comparison of WebXR and commercial game-engine approaches across key dimensions relevant to Metaverse development. Scores represent relative strengths discussed throughout the paper and are intended as an interpretative synthesis rather than a quantitative benchmark.}
%\label{fig:radar_chart}
%\end{figure}

\subsection{WebXR Architecture and Capability Abstraction}

To understand the differences between browser-based and native XR development, it is important to distinguish between the underlying standards that govern hardware access. A widely adopted foundation for native XR interoperability is OpenXR\footnote{\url{https://www.khronos.org/openxr/}}, an open, cross-vendor specification managed by the Khronos Group. OpenXR acts as a device abstraction layer that allows native applications and commercial game engines to interact directly with low-level hardware components across different headsets, including Meta Quest, Apple Vision Pro, HTC Vive, and others.

WebXR adopts a different architectural philosophy. Rather than replacing OpenXR, the WebXR Device API typically operates above native XR runtimes through the browser. The execution pipeline flows from the physical hardware to the native OpenXR implementation, which exposes device information to the browser. The browser subsequently filters and abstracts these capabilities before exposing them through standardized WebXR interfaces. Finally, higher-level libraries such as Babylon.js, Three.js, and A-Frame provide additional abstractions that simplify application development.

This layered architecture offers important advantages in terms of portability, interoperability, and security, but also introduces a degree of abstraction between applications and the underlying hardware. Consequently, browser-based applications generally gain access to new device capabilities only after corresponding browser implementations and W3C standardization efforts become available. Native applications, by contrast, can often exploit vendor-specific OpenXR extensions immediately after they are released.

The WebXR ecosystem itself is organized as a collection of modular W3C specifications rather than a monolithic framework. Different modules progressively extend browser capabilities in areas such as environmental understanding, interaction, input tracking, and rendering. This modular design allows browser vendors to implement new functionality incrementally while maintaining compatibility with existing applications.

\begin{table*}[t]
\caption{OVERVIEW OF SELECTED WEBXR SPECIFICATIONS AND OBSERVED IMPLEMENTATION CHARACTERISTICS (Status as of 2026)}
\label{tab:webxr_features}
\centering
\footnotesize 
\renewcommand{\arraystretch}{0.85} 
\setlength{\tabcolsep}{3.5pt}      
\begin{tabularx}{\textwidth}{>{\raggedright\arraybackslash}p{2.0cm}|l|L|l|L}
\toprule
\textbf{Category} & \textbf{W3C Specification Module} & \textbf{Target Capabilities} & \textbf{Browser Implementation} & \textbf{Privacy Sensitivity} \\ 
\midrule
\rowcolor{myblue}
\textbf{Real-World} & WebXR Hit Test API & Spatial raycasting against physical geometry & Broad support (Chromium, Safari) & Limited (Raw intersection vectors) \\
\rowcolor{myblue}
\textbf{Interaction} & WebXR Anchors & Persistent spatial positioning markers & Broad support (Meta, Apple, Pico) & Limited (OS-managed) \\ 
\midrule
& WebXR Plane Detection API & Identification of flat surfaces (floors, walls) & Wide support (Chromium) & Moderate (Room layout exposure) \\
\textbf{Environmental} & WebXR Mesh Detection API & Dense triangle geometric mesh extraction & Experimental (Behind flags) & Elevated (Fingerprinting potential) \\
\textbf{Perception} & WebXR Light Estimation API & Ambient color, intensity, and cube-map reflection & Broad support (Mobile AR, XR browsers) & Limited (Lighting coefficients) \\ 
\midrule
\rowcolor{myblue}
\textbf{Input \&} & WebXR Hand Input API & 25-joint skeletal finger and joint tracking & Broad support (Quest, Vision Pro) & Moderate (Biomechanical traits) \\
\rowcolor{myblue}
\textbf{Tracking} & WebXR Gamepads Module & Mapping buttons and axes of hardware controllers & Broad support (Core specification) & Minimal (Permission-based) \\
\rowcolor{myblue}
& Transient Input Model & Screen-space gaze tracking and pinch events & Broad support (VisionOS / Safari) & Limited (Selection intent only) \\ 
\midrule
\textbf{Visuals \&} & WebXR Layers API & Direct texture composition bypass to hardware & Wide support & Minimal (Rendering only) \\
\textbf{Rendering} & WebXR Multi-View Extension & Single-pass stereoscopic GPU draw-call loops & Multi-runtime support (WebGL2 / WebGPU) & Minimal (GPU optimization) \\ 
\bottomrule
\end{tabularx}
\end{table*}

Table~\ref{tab:webxr_features} summarizes representative WebXR-related specifications relevant to Metaverse development. The table should be interpreted as an illustrative overview derived from W3C specifications, browser documentation, and publicly available implementation reports rather than as an exhaustive or permanent classification, since browser support and specification maturity continue to evolve.

One important consequence of this architecture is the balance between hardware access and user privacy. Native applications generally receive direct access to spatial information exposed by the underlying runtime, whereas browser-based applications expose only standardized abstractions designed to reduce privacy risks. For example, the \textit{WebXR Hit Test API} enables virtual objects to interact with physical surfaces without exposing detailed information about the surrounding environment. More advanced capabilities, such as plane detection or environmental meshing, require explicit user permissions and remain subject to browser-specific implementation decisions.

Finally, the practical impact of browser fragmentation is partially mitigated by the surrounding open-source ecosystem. Frameworks such as Babylon.js and Three.js encapsulate many browser-specific implementation details, allowing developers to target multiple browsers and devices through a common programming interface while benefiting from progressively evolving WebXR capabilities.

\subsection{Hardware Access, Device Independence, and Platform Support}
A major promise of WebXR is device independence. A WebXR application can, in principle, run across headsets, desktop browsers, smartphones, and tablets, provided that the browser and device expose the required APIs. This is particularly attractive for institutions that cannot control which headset a student, visitor, or collaborator owns. It also reduces the need to compile and maintain separate builds for Quest, Android XR, Apple Vision Pro, desktop VR, or mobile AR.

Commercial engines, by contrast, offer stronger integration with specific hardware. Native applications can often access advanced features earlier than WebXR, including optimized hand tracking, passthrough, scene reconstruction, eye tracking, foveated rendering, spatial anchors, platform avatars, and vendor-specific interaction models. For research prototypes that require experimental hardware features, or for industrial applications where performance and sensing precision are critical, native engines remain advantageous.

The difference can be summarized as follows: WebXR prioritizes reach; native engines prioritize depth. WebXR is preferable when the main requirement is that many users can enter an experience easily. Native engines are preferable when the application depends on the most advanced capabilities of a specific headset.
This assessment represents merely a snapshot of the contemporary landscape rather than a permanent architectural boundary. A historical analogy can be drawn with web productivity tools, whose capabilities expanded substantially over time. While such trajectories are not guaranteed to repeat in XR, they suggest that browser-based limitations should be interpreted cautiously.

\subsection{Software Architecture and Interoperability}
The Metaverse requires interoperability at multiple levels: identity, avatars, assets, networking, payments, moderation, and persistence. WebXR does not solve all of these problems, but it fits naturally within the Web's existing interoperability model. As discussed previously, URLs, HTTP(S), WebRTC, OAuth, WebAssembly, WebGPU, glTF, and WebAudio can be combined to build modular systems. This allows XR experiences to reuse the Web's mature infrastructure rather than reinventing distribution, linking, and access control.

Recent work has highlighted that interoperability challenges extend beyond avatars, assets, and networking to include audio technologies. Boem et al.~\cite{boem2025issues} argue that the absence of common standards for spatial audio rendering, audio processing, and synchronization across immersive platforms represents a significant obstacle to the realization of interoperable Metaverse ecosystems. Their analysis suggests that standardization efforts are required not only at the visual level but also for multimodal components that support communication, collaboration, and creative expression.

Commercial engines can also support open formats and network protocols, but engine projects often become large binary applications tied to specific versions, plugins, and build targets. This can create lock-in. A Unity or Unreal project may be portable in theory, but in practice it can depend on proprietary packages, SDK versions, rendering pipelines, and app-store requirements. WebXR applications can also suffer from dependencies, especially on JavaScript frameworks, but the Web's backward-compatibility culture provides a stronger long-term preservation model than many native XR stacks.

It is worth noting that commercial game engines like Unity are built on a multi-target compilation paradigm, allowing developers to deploy a single codebase across native platforms such as mobile Android, iOS, and Windows, \textit{as well as the web}. To achieve web deployment, the engine compiles its native C\# codebase into JavaScript and WebAssembly, while utilizing intermediate wrappers to take advantage of specific browser APIs, including WebXR, to some extent. However, this workflow currently remains experimental and structurally heavy compared to native web architectures. A more comprehensive analysis of \textit{this hybrid pipeline}, its performance trade-offs, and its limitations is provided in Section \ref{ref:audioPerf}.

\subsection{Deployment, Maintenance, and Updates}

Beyond the architectural differences discussed previously, the two ecosystems differ substantially in their deployment and maintenance workflows. Native XR applications generally require platform-specific packaging, application installation, app-store submission or sideloading, and periodic distribution of updated binaries. By contrast, WebXR applications are deployed through standard web infrastructure and are immediately accessible through a compatible browser, allowing updates to be performed centrally on the server without requiring user intervention.

This difference is particularly relevant in educational, research, and public institutional settings. When XR headsets are shared among multiple users, native deployment workflows often require account management, device pairing, application installation, firmware compatibility checks, and repeated device resets. Browser-based deployment can significantly reduce this operational burden, since users only require access to a compatible browser and a network connection.

The URL-centric nature of WebXR also simplifies session orchestration for collaborative applications. In native ecosystems, establishing a multi-user session may require participants to install identical application versions, exchange room identifiers, or rely on platform-specific social infrastructures. Within WebXR, collaborative sessions can instead be initiated through standard hyperlinks containing the required session parameters, facilitating rapid onboarding and reducing friction for first-time users.

Commercial app deployment has benefits too. App stores provide visibility, payment infrastructure, parental controls, age ratings, malware checks, and user-review mechanisms. Apple's App Review Guidelines, for example, explicitly define review expectations concerning safety, content, privacy, and app behavior\footnote{\url{https://developer.apple.com/app-store/review/guidelines/}}. These mechanisms are imperfect and may be restrictive, but they provide a governance layer that the open Web does not automatically supply.
Therefore, WebXR increases freedom but also transfers responsibility. Developers who bypass app stores gain flexibility, but must implement their own safety, privacy, moderation, analytics governance, and child-protection practices.

\subsection{Economic and Market Considerations}
Commercial platforms create costs and dependencies. Apple Developer Program membership currently costs \$99 per year; Apple also reviews applications before distribution and can reject applications under its guidelines. App-store ecosystems may also impose restrictions on monetization, external payments, subscriptions, and digital goods. In the European Union, the Digital Markets Act has pressured gatekeepers to modify their rules, but recent changes have introduced complex fee structures rather than fully eliminating platform dependency.%\footnote{\url{https://www.theverge.com/news/693512/apple-eu-dma-app-store-concessions}}.

%WebXR offers a route around some of these constraints. A developer can host an XR experience independently, monetize through web-based payments, and avoid app-store review. This is particularly relevant for European institutions and small developers seeking autonomy from US-controlled distribution channels. However, WebXR does not eliminate all platform power. Browser vendors, headset manufacturers, search engines, payment processors, cloud providers, and certificate authorities still shape what is technically and economically possible.

While the centralized, store-based model is easy to comprehend despite its associated overhead and platform taxes, web-based monetization offers a  more open, diverse, and flexible landscape. Because WebXR environments are decoupled from proprietary marketplaces, developers can implement a wide array of alternative commercial architectures tailored to their specific needs. These include standard web-based subscriptions managed through direct processing gateways (e.g., Stripe, PayPal), consumption-based metered billing, paywalled premium access layers, embedded contextual web advertising, or emerging Web Monetization standards that enable continuous micro-payments via streaming open protocols. By bypassing the mandatory 15\% to 30\% platform fee typically extracted by native app stores, small-to-medium enterprises and independent creators can retain a significantly higher share of their revenue.

Conversely, for non-commercial or public-interest deployment models, the web platform may provide practical advantages over proprietary marketplaces. Educational resources, civic software, public sector tools, and academic research prototypes often operate entirely outside the paradigm of financial monetization. Forcing these open-access initiatives into commercial app stores mandates compliance with restrictive administrative guidelines, requires ongoing platform subscription maintenance fees, and introduces unnecessary gatekeeping for end-users. By utilizing standard web hosting, these public-interest frameworks can distribute immersive experiences freely and universally, ensuring that access is gated by neither app-store algorithms nor economic paywalls.

WebXR thus offers a route around some of these constraints. A developer can host an XR experience independently, monetize through web-based payments, and avoid app-store review. This is particularly relevant for European institutions and small developers seeking autonomy from US-controlled distribution channels. However, WebXR does not eliminate all platform power. Browser vendors, headset manufacturers, search engines, payment processors, cloud providers, and certificate authorities still shape what is technically and economically possible.

\subsection{Performance and Fidelity}
Commercial game engines generally provide stronger performance for computationally demanding XR experiences. Native engines offer mature rendering pipelines, platform-specific optimization, profiling tools, asset streaming, and direct integration with vendor SDKs. Unreal Engine is particularly associated with high-fidelity rendering, while Unity remains widely adopted because of its broad XR ecosystem support.

WebXR performance has improved substantially through technologies such as WebAssembly and WebGPU, enabling browser-based XR experiences with increasing graphical and computational sophistication. For many applications --- including education, collaborative environments, data visualization, museums, telepresence, and lightweight immersive experiences --- WebXR may provide sufficient performance and perceptual fidelity. However, browser execution environments still introduce limitations related to sandboxing, memory management, heterogeneous API support, and constrained access to device-specific optimizations.

These observations are consistent with prior work reporting that browser-based XR environments are often perceived as less performant than native applications for computationally demanding experiences \cite{zubair2021long}. At the same time, browser-based runtimes continue to evolve, and recent developments suggest that some historical limitations have narrowed. For example, WebAssembly-based integration of mature software components such as physics engines has enabled increasingly sophisticated interactive XR simulations in browser environments~\cite{buffa2025embodied}.

%The introduction of WebAssembly (WASM), for instance, has substantially reduced several historical limitations by bringing heavy computational architectures to the web platform that were previously restricted to native execution layers. A prime example is the porting of the Havok Physics engine—a production-proven, industry-standard tool traditionally powering native AAA video games—directly into the browser runtime via integration with the Babylon.js library. This native-grade physics capability enables complex, low-latency WebXR simulations. In collaborative, real-time virtual music-making environments, for example, developers used Havok in virtual drum kits that map highly responsive physical interactions on the fly \cite{buffa2025embodied}. In this experiment, the engine accurately simulates complex collision physics, such as the direct tactile response of a stick hitting a drum skin or the realistic, secondary wobbling movements of suspended cymbals during gameplay. This demonstrates that performance constraints are rapidly shifting from an architectural impossibility to an implementation detail.

\subsection{Audio and Haptic Capabilities}
\label{ref:audioPerf}
The comparison between WebXR and commercial game engines extends beyond graphics and rendering performance to  other key modalities of immersive experiences, notably spatial audio and haptic interaction.

Spatial audio is a relatively mature component of the Web ecosystem through the Web Audio API, which supports real-time audio synthesis, filtering, spatialization, and integration with networked communication systems \cite{tomasetti2023spatial,boem2025spatial}. Combined with technologies such as WebRTC, browser-based XR environments can support positional audio, collaborative interaction, and shared immersive experiences \cite{dziwis2023orchestra, buffa2025interactive, boem2023musical}. Prior work has demonstrated the feasibility of implementing immersive and collaborative sonic environments directly within WebXR applications \cite{tomasetti2023spatial, boem2023musical}%\footnote{Several of the cited empirical studies  originate from the authors' own prior work, reflecting the scarcity of independent evaluations in this emerging area rather than selective citation practice}.

Commercial game engines nevertheless retain advantages for demanding audio production workflows. Platforms such as Unity and Unreal support mature middleware ecosystems (e.g., FMOD, Wwise), advanced acoustic simulation, and fine-grained control over audio processing, which remain important for large-scale or professionally engineered XR experiences. At the same time, recent developments suggest that the gap between native and browser-based audio workflows is narrowing. WebAssembly-based approaches and Web Audio Modules increasingly support sophisticated browser-native signal-processing pipelines, while recent demonstrations have shown that professional audio middleware can increasingly be adapted to browser-oriented environments \cite{MilotWwiseDemoWAC2022, MilotWwiseWAC2022Video, buffa2025ten, buffa2025interactive}.

The situation differs more substantially for haptics. Current WebXR implementations provide only limited and heterogeneous support for haptic feedback, typically restricted to simple controller vibrations when supported by the browser and underlying hardware. 
%\footnote{\url{https://lochie-web-haptics-50.mintlify.app/introduction}}. 
Access to advanced haptic devices, wearable tactile interfaces, force-feedback systems, and experimental research hardware remains limited within browser-based environments.
By contrast, commercial game engines provide broader access to vendor SDKs and specialized haptic devices, making native approaches preferable for applications in which tactile interaction constitutes a central component of the experience.

Overall, the comparison suggests that WebXR is already suitable for many audio-centric Metaverse scenarios, including social XR, collaborative creativity, and virtual performances, whereas native engines remain preferable when sophisticated haptic interaction or low-level hardware control is required.

\subsection{Developer Accessibility}
WebXR may lower  entry barriers for developers already familiar with web technologies. HTML, JavaScript, TypeScript, CSS, Three.js, Babylon.js, and A-Frame are more accessible to many students and creative technologists than full native-engine pipelines. WebXR also supports rapid iteration: changes can be deployed immediately to a server rather than compiled and submitted to a store.
At the same time, empirical work has identified challenges associated with WebXR development, including fragmented documentation, evolving standards, limited learning resources, and heterogeneous browser support\cite{zubair2021long}. Consequently, WebXR is not inherently less complex, but rather better aligned with teams possessing web-development expertise.

Commercial game engines provide highly integrated authoring environments that may be more accessible to teams working within established 3D production workflows. Visual editors, asset pipelines, animation systems, profiling tools, and XR interaction frameworks can reduce implementation complexity for designers, artists, and multidisciplinary production teams. As a result, native engines may remain preferable for workflows centered on mature 3D content production.

\subsection{Ethics, Governance, and Safety}
A comparison concerning the ethical aspects represents a complex matter. WebXR supports openness, access, and autonomy, but weakens centralized review. Commercial stores can enforce safety checks, age ratings, content moderation, privacy disclosures, and payment protections. For children and vulnerable users in general, such mechanisms are valuable.
At the same time, centralized control raises other ethical issues: arbitrary rejection, opaque governance, unequal market access, surveillance, platform lock-in, and dependency on corporate policies. XR devices are especially sensitive because they can collect spatial maps, body movement, gaze, voice, interaction behavior, and biometric-adjacent data. A closed platform can centralize these data flows, while an open WebXR ecosystem can diversify control but may also fragment accountability.

Therefore, a responsible WebXR Metaverse requires explicit governance mechanisms: privacy-by-design, transparent data collection, safe defaults, content moderation, age-appropriate design, accessibility testing, and open auditing. It is worth emphasizing that openness alone does not guarantee ethical behavior. However, open technologies can facilitate transparency, independent auditing, and public scrutiny, thereby creating conditions that support accountability and responsible innovation.

\subsection{Sustainability and Obsolescence}
%Sustainability includes energy use, hardware lifetime, software maintainability, and institutional continuity. Native XR ecosystems can accelerate obsolescence when devices lose store support or security updates. For example, Meta ended Quest 1 security updates and bug fixes after August 31, 2024, while new Quest 1-exclusive apps had already been stopped earlier in 2024. Such lifecycle decisions can make otherwise functional hardware less viable for schools, labs, and cultural institutions.

Sustainability encompasses energy use, hardware lifetime, software maintainability, and institutional continuity. Native XR ecosystems frequently accelerate hardware obsolescence through vendor-enforced lifecycles, such as when devices abruptly lose marketplace access or critical security updates. For instance, Meta officially terminated security updates and bug fixes for the Quest 1 after August 31, 2024, having already blocked new app deployments earlier that year. Such platform-governance decisions prematurely devalue perfectly functional hardware assets, rendering them unviable for budget-constrained schools, research laboratories, and cultural institutions.

%WebXR can mitigate some forms of obsolescence because content is not tied to a single app store or binary package. If a device has a compatible browser, a web experience may continue to function. The Web's long-standing backward-compatibility ethos is an important sustainability asset. Nevertheless, WebXR is not immune to obsolescence: browsers remove features, APIs evolve, certificates expire, frameworks become unmaintained, and hardware still ages. The sustainability advantage of WebXR is therefore relative, not absolute.

WebXR may mitigate these forms of platform lock-in because browser-delivered experiences are decoupled from specific binary deployment packages and centralized app stores. If an aging device maintains a compatible browser, the immersive application can remain functional, leveraging the web's foundational ethos of long-term backward compatibility. Nevertheless, WebXR is not entirely immune to obsolescence: security certificates expire, third-party frameworks fall out of maintenance, and the underlying physical hardware eventually degrades. The sustainability advantage of the web is therefore relative rather than absolute.

%However, sustainability challenges also exist within WebXR ecosystems. Interviews conducted by Zubair and Anyameluhor \cite{zubair2021long} revealed that developers frequently struggle to keep projects operational as browsers, APIs, and frameworks evolve over time. Consequently, while WebXR may mitigate certain forms of platform obsolescence associated with proprietary app stores and device ecosystems, it does not eliminate the need for continuous software maintenance.

The operational maintenance burden within WebXR varies significantly based on the maturity level of the underlying browser specifications. Sustainability challenges are most pronounced when developers build applications using cutting-edge features that are still in their early technical track stages (such as working drafts). This dynamic aligns with the empirical findings (2021) of Zubair and Anyameluhor \cite{zubair2021long}, whose developer interviews revealed frequent struggles to keep projects operational as volatile experimental frameworks and APIs evolved beneath them. 

However, this systemic friction is temporary and directly tied to the W3C standardization lifecycle. For example, while the core WebXR Device API is currently (2026) a Candidate Recommendation Draft
%\footnote{\url{https://www.w3.org/TR/webxr/}}
, meaning it is highly stable but technically still a work in progress, adjacent specifications like the Web Audio API and Web MIDI API have achieved full W3C Recommendation status in 2021, operating as frozen, finalized standards. Once an API advances to a formal W3C Recommendation or RFC equivalent, browser vendors generally maintains long-term backward compatibility, substantially reducing  maintenance burden. Furthermore, because the structural and syntactic deltas between mature candidate drafts and finalized specifications are historically minor, developers may reasonably invest in early-stage WebXR implementations with reduced risk of disruptive code obsolescence.

%% 

% Maybe we can also talk about Mozilla Hub which was closed???

%
%To synthesize the findings discussed throughout this section, Figure~\ref{fig:radar_comparison} summarizes the relative strengths and limitations of WebXR and commercial game-engine approaches across the main dimensions considered in this study.
%
%Table~\ref{tab:comparison} summarizes the main differences between WebXR and commercial game-engine approaches across the principal dimensions considered in this paper. 
%Figure~\ref{fig:radar_comparison} provides a visual synthesis of the trade-offs between WebXR and commercial game-engine approaches across the principal dimensions discussed in this paper. The ratings are intended to be qualitative rather than quantitative and summarize the overall trends emerging from the comparative analysis. Rather than identifying a universally superior approach, the figure highlights the complementary strengths of the two paradigms and the multidimensional nature of technology selection for Metaverse development.
%

\section{Discussion}

\begin{figure*}[ht!]
\centerline{\includegraphics[width=\textwidth]{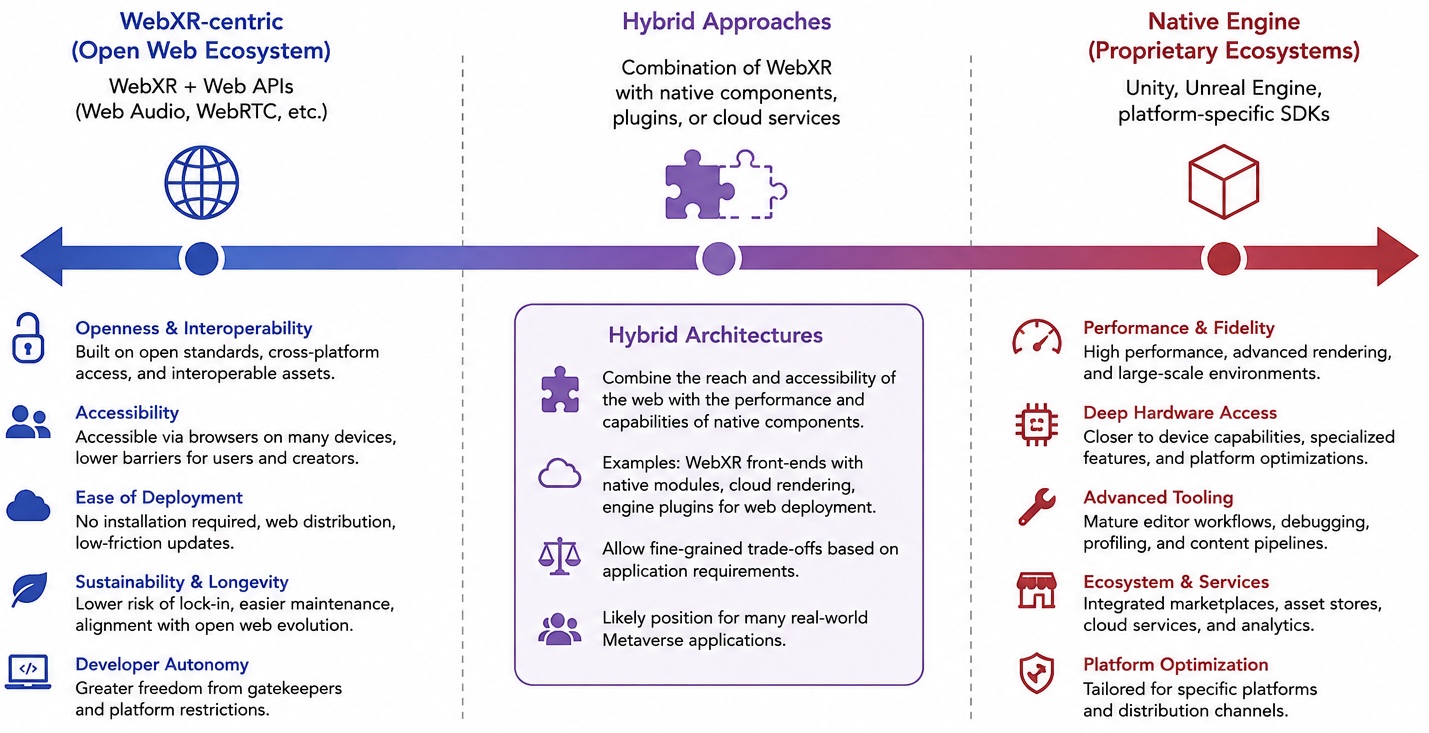}}
\caption{Conceptual continuum of Metaverse development approaches. WebXR-based solutions and commercial game-engine ecosystems represent different positions along a spectrum of socio-technical trade-offs. While WebXR tends to emphasize openness, interoperability, accessibility, and deployment flexibility, native-engine approaches prioritize performance, advanced tooling, and deep hardware integration. Many practical Metaverse applications are likely to adopt hybrid architectures occupying intermediate positions on the continuum.}
\label{fig:continuum}
\end{figure*}

The analysis presented in this paper allows us to answer the three research questions introduced in Sec.~\ref{sec:intro}. A central finding emerging from the comparative analysis is that WebXR and commercial game-engine ecosystems should not be interpreted as mutually exclusive alternatives. Rather, they occupy different positions along a continuum of socio-technical trade-offs encompassing technical, organizational, economic, and governance dimensions (see Fig.~\ref{fig:continuum}). At one end of this continuum lie solutions that prioritize openness, interoperability, accessibility, ease of deployment, and long-term maintainability; at the other end lie solutions that prioritize performance, deep hardware integration, advanced tooling, and platform-specific optimization. Consequently, the choice between the two approaches should be driven not only by performance requirements, but also by broader considerations related to sustainability, accessibility, developer autonomy, governance, and the intended application domain. Most real-world Metaverse applications are likely to occupy intermediate positions on this continuum and, therefore, require balancing these competing objectives rather than selecting one paradigm unconditionally.

\subsection{RQ1: Technical Advantages and Limitations of WebXR}

Our analysis indicates that WebXR's principal technical advantage lies in its browser-based deployment model. Experiences can be accessed through a URL without requiring installation, significantly reducing friction for end users and simplifying deployment in educational, research, and public-facing contexts. Furthermore, WebXR benefits from the interoperability of the broader web ecosystem, enabling integration with established technologies such as WebRTC, WebAudio, WebAssembly, WebGPU, and open asset formats.

Another important advantage is platform independence. Rather than maintaining multiple native builds for different XR devices, developers can target a common web-based execution environment. This facilitates broader reach and can reduce development and maintenance effort, particularly for projects with limited resources.

However, WebXR still presents limitations. Browser-based execution introduces constraints in terms of performance optimization, memory management, and access to advanced device-specific capabilities. Although recent developments such as WebGPU and WebAssembly have considerably narrowed the performance gap, native applications developed with commercial engines generally retain advantages for graphically intensive, latency-sensitive, or highly optimized XR experiences. Similarly, support for emerging hardware features often becomes available earlier through native SDKs than through browser APIs.

Overall, the results suggest that WebXR is technically mature enough for a large class of Metaverse applications, including education, collaboration, social interaction, cultural heritage, and lightweight immersive experiences, while native game engines remain preferable for applications requiring maximum performance and deep hardware integration.

An important observation concerns multimodal interaction. The technical gap between WebXR and native applications is not uniform across modalities. In the case of spatial audio, browser-based technologies have matured considerably and can support many social, collaborative, and creative XR experiences with high perceptual quality \cite{boem2025spatial}. Consequently, WebXR is already a viable platform for numerous audio-centric Metaverse applications \cite{buffa2024using}. By contrast, support for advanced haptic interaction remains substantially more limited, as browser environments typically provide only restricted access to specialized haptic hardware and low-level device interfaces. This asymmetry suggests that the trade-off between accessibility and technical depth is not uniform across all XR modalities. While browser-based limitations remain more visible for computationally demanding rendering pipelines and advanced visual fidelity, developments such as WebAssembly have reduced several historical constraints affecting browser-based computation. Mature software components, including physics engines and advanced audio-processing workflows implemented through Web Audio Modules can increasingly operate within browser environments, supporting richer interactive and perceptual experiences in non-visual channels. Consequently, the suitability of WebXR depends not only on graphical requirements but also on the sensory modalities emphasized by a given application.

\subsection{RQ2: Interoperability, Governance, Developer Autonomy, and Sustainability}

Beyond the technical dimensions examined above, the governance and sustainability implications of the two approaches may ultimately be more consequential for institutions building long-term Metaverse infrastructure. Commercial XR ecosystems are frequently characterized by multiple layers of dependency, including engine vendors, platform providers, app-store operators, device manufacturers, and account-management systems. While these actors provide valuable services such as distribution infrastructure, security reviews, payment systems, and developer support, they also introduce forms of platform dependence that can affect long-term project viability and developer autonomy.

WebXR, in contrast, inherits the governance model of the Web, which is based on open standards and decentralized deployment. Developers can publish experiences independently, update them without app-store approval processes, and distribute them directly through URLs. This reduces barriers to entry and can facilitate innovation, experimentation, and broader participation in Metaverse development.

From a sustainability perspective, WebXR also presents notable advantages. Browser-based deployment reduces the need for application-specific installation and maintenance procedures and may mitigate some forms of software obsolescence associated with discontinued app stores, deprecated SDKs, or abandoned device ecosystems. Although WebXR remains dependent on browser vendors and evolving standards, the long history of backward compatibility within web technologies offers a stronger foundation for long-term preservation than many proprietary XR ecosystems.

At the same time, openness should not be confused with maturity. Empirical evidence from WebXR practitioners suggests that the benefits of browser-based deployment are often accompanied by challenges related to fragmented browser support, evolving standards, device heterogeneity, and long-term maintenance requirements~\cite{zubair2021long}. Therefore, although WebXR aligns closely with the vision of an open Metaverse, achieving this vision in practice requires continued standardization efforts and improved tooling for developers.

These findings suggest that WebXR aligns more closely with visions of an open and interoperable Metaverse, while commercial ecosystems offer greater centralization, tighter integration, and more controlled user experiences.

\subsection{RQ3: Appropriate Application Domains for Each Approach}

The answer emerging from this analysis is not binary. Instead, the comparison suggests the existence of a continuum of design choices between highly open, browser-based approaches and highly optimized, platform-specific native solutions.
WebXR is especially advantageous when accessibility, interoperability, ease of deployment, and institutional sustainability are primary objectives. Examples include educational platforms, virtual museums, scientific visualizations, collaborative workspaces, social XR environments, public-sector applications, and research prototypes. In these contexts, the ability to access immersive experiences through a browser can significantly increase participation while reducing operational complexity.
Commercial game engines remain the preferred solution for applications requiring state-of-the-art graphics, advanced rendering pipelines, low-latency interaction, extensive use of platform-specific SDKs, or highly optimized execution. Large-scale games, industrial simulations, engineering digital twins, military training systems, and professional XR products often fall into this category.

Importantly, the analysis also suggests the emergence of hybrid approaches. As browser technologies continue to evolve and native engines increasingly support web deployment targets, future Metaverse ecosystems may combine the strengths of both paradigms. In such scenarios, WebXR may provide interoperable access layers and onboarding mechanisms, while native engines may power the most computationally demanding components.

\subsection{Implications for the Future Metaverse}

Beyond the specific comparison between WebXR and commercial engines, the results have broader implications for the future evolution of the Metaverse. If the Metaverse is envisioned as an open ecosystem rather than a collection of disconnected proprietary platforms, then interoperability, portability, and governance become as important as rendering performance or hardware capabilities.
In this regard, WebXR should not be viewed merely as a technical framework for browser-based XR applications. Rather, it represents an architectural approach that aligns with principles of openness, decentralization, and technological sovereignty. This perspective may be particularly relevant for educational institutions, public administrations, cultural organizations, and research communities seeking to avoid excessive dependence on proprietary ecosystems.

At the same time, openness alone does not guarantee safety, accessibility, or ethical behavior. The reduced gatekeeping associated with WebXR places greater responsibility on developers and organizations to implement appropriate security, privacy, moderation, and accessibility mechanisms. Consequently, the future success of WebXR-based Metaverse ecosystems will depend not only on technical maturity but also on the development of robust governance and trust frameworks.

Another important implication of the analysis is that future Metaverse ecosystems are unlikely to converge toward a single dominant technological paradigm. Rather, the evidence suggests the emergence of hybrid architectures that combine the openness and accessibility of WebXR with the performance and hardware capabilities of native engines. WebXR may increasingly serve as an interoperable access layer for onboarding, education, public-facing experiences, and lightweight collaborative environments, while native engines may continue to power computationally demanding applications requiring advanced rendering, simulation, or specialized hardware integration. As technologies such as WebAssembly, WebGPU, and browser-native access to XR capabilities continue to mature, the boundary between web-based and native XR development is likely to become progressively more porous.

\section{Scope and limitations}

As a position paper, this work does not aim to provide controlled benchmarking experiments or exhaustive quantitative evaluation. Instead, it offers an evidence-informed synthesis of technical documentation, standards specifications, prior empirical findings, and documented industrial practices. Consequently, the observations presented should be interpreted as a socio-technical perspective on evolving XR ecosystems rather than definitive measurements of technical superiority. Furthermore, because browser implementations, XR hardware capabilities, and standards continue to evolve rapidly, some conclusions may vary over time and across platforms.
Furthermore, the coverage of audio and haptic technologies reflects the current state of available literature in these subfields, where peer-reviewed empirical work remains sparse and some foundational contributions originate from the authors' research group.
\section{Conclusions} 

This paper compared open WebXR technologies with conventional commercial game-engine pipelines for Metaverse development. WebXR provides important advantages in accessibility, interoperability, deployment, institutional autonomy, and sustainability. It allows immersive experiences to be distributed through links rather than app stores, reduces friction for shared devices and educational contexts, and aligns with the vision of an open Metaverse based on standards rather than closed ecosystems.

However, WebXR is not a complete replacement for commercial engines. Native engines remain superior for many high-performance, graphically intensive, and hardware-specific XR applications. They also provide mature tooling, store-based governance, monetization infrastructure, and production workflows that WebXR cannot yet fully match.
Moreover, while WebXR is already sufficiently mature for many social, educational, collaborative, and audio-oriented Metaverse experiences, it still lags behind native engines in support for advanced haptics, low-level hardware access, and platform-specific optimizations.

The central conclusion is therefore strategic rather than ideological. Rather than representing mutually exclusive alternatives, WebXR and commercial game engines should be understood as occupying different positions along a continuum of socio-technical trade-offs. At one end of this continuum lie solutions emphasizing openness, interoperability, accessibility, deployment flexibility, and long-term sustainability; at the other end lie solutions emphasizing performance, advanced tooling, and deep hardware integration. Consequently, the choice between the two approaches should be guided by the requirements and priorities of the intended application rather than by a universal preference for one ecosystem over the other. For the Metaverse to remain accessible, ethical, and sustainable, future XR ecosystems should support both paths while strengthening open standards, transparent governance, and long-term interoperability.

%%%%%%%%%%%%%%%%%%%%%%%%%%%%%%%%%%%%%%%%%%%%%%%%%%%%%%%%%%%%%%%%%%%%%%%%%%%%%%%%%%%%%%%%%%%%%%%%%%

\bibliographystyle{IEEEtran}
\bibliography{article_specific.bib,IoMusT.bib,musical_XR_and_MM.bib,NMP.bib,Turchet.bib,WebAudio}

\end{document}